\documentclass[twoside,slac_one]{revtex4}
\usepackage{graphicx}
\usepackage{fancyhdr}
\usepackage{amsmath} % American Mathematics Society standards
\usepackage{bm}% bold math
\usepackage{amsxtra}
\usepackage{amssymb}
\usepackage{amsthm}
\usepackage{latexsym}
\usepackage{lscape}

\begin{document}

\title{Probing Hot and Dense Nuclear Matter with Particle Correlations and Jets at RHIC}


\author{H. Pei}
\affiliation{Department of Physics and Astronomy, University of Illinois at Chicago, Chicago, IL, USA}

\begin{abstract}
The hot and dense medium created at RHIC, called Quark and Gluon Plasma (QGP) has been a hot topic in the last ten years. Due to the high multiplicities in such heavy-ion collision events, particle correlations using either trigger particles, or fully-reconstructed jets, become not only useful but necessary, in addition to the single particle observables. In this paper the most recent work studying this medium will be shown, including both on bulk properties and tagged events. 

\end{abstract}

\maketitle

\thispagestyle{fancy}


%%%%%%%%%%%%%%%%%%%%%%%%%%%%%%%%%%
\section{Introduction}

The central goal of RHIC/LHC heavy-ion program is to study the quantitative properties of the phases of QCD. The bulk properties of this hot and dense QGP medium is going to be used to extract the physics characters: $T, C_{s}, \hat{q}, \eta, \zeta,$ etc. While the single particles observables are the natural points, the following sections will show the methods of correlations are also powerful and necessary.

%%%%%%%%%%%%%%%%%%%%%%%%%%%%%%%%%%
\section{Beginning from singles $R_{AA}$}

The single particles $R_{AA}$ measurement at RHIC shows clearly PID dependence. The mesons, whether of light quarks or charm/bottom quarks, all indicate strong suppression patterns at high-$p_{T}$. ~\cite{phenix_pid_raa, star_pid_raa, phenix_npe_raa}. On the other hand, the ``baryon anomaly'' and direct-$\gamma$ $R_{AA}$ measurements show strong medium effects on the particle production mechanisms. ~\cite{phenix_pid_raa, star_pid_raa, phenix_gamma_raa}.
This medium effects is further studied in the recent LHC data, where the $\sqrt{s_{NN}}$ is more than a factor of 10 higher. The $R_{AA}$ of mesons is reported to be very close at LHC as RHIC at 5<$p_{T}$<20 GeV/$c$. ~\cite{alice_pid_raa} It has been discussed if the same ``Quark soup'' had been cooked at LHC and RHIC. 
Thus, it is necessary to introduce correlation method to study the medium in further details.

%%%%%%%%%%%%%%%%%%%%%%%%%%%%%%%%%%
\section{Correlations}

Correlations can be studied in two ways, triggered and untriggered. The triggers are usually high-$p_{T}$ particles as proxies of jets, or jets themselves. Because jets are considered to originate from hard-scattering of partons which happen at the early age of QGP formation, they have a high chance of carrying the information of medium by flying through and interact with medium, thus are good probes of medium.
However, the RAW correlations contain not only jets, but also bulk medium information, mainly the flow items ($v_{n}$), and these properties have their own centrality dependence other than that of jets.
These flow factors then bring complexity to the correlations study. For example, RHIC has reported the ``cone'' and ``ridge'' structures ~\cite{phenix_ppg_074, star_ridge}. While the first even order flow $v_{2}$ has been subtracted, questions still rise as whether these structures are evidence of modified jets production, or these are fluctuations and/or evolution of medium itself that coincide with trigger particles.

Therefore, it is necessary to disentangle these flow factors from jets correlations at much detail. Higher order $v_{n}$, and not only even order but also odd orders, needs to be considered and subtracted carefully. This is where the untriggered correlations come to use, which will bring us the so-called $v_n (2)$, $v_n (4)$, etc.

%%%%%%%%%%%%%%%%%%%%%%%%%%%%%%%%%%
\subsection{Before $v_{3}$ era}

While the flow items can be measured through untriggered correlations, the effort to disentangle them from jets were already made even earlier through multi-particle correlations. In the $\Delta\eta-\Delta\eta$ correlation paper ~\cite{star_deta_deta}, by studying the cross-pair densities at ``jet'' and ``ridge'' regions, the possibility of particles correlated in physics between these two was found to be close to zero within errors. This paper then claimed ``No correlation is found between production of the ridge and production of the jet-like particles, suggesting the ridge may be formed from the bulk medium itself.''

%%%%%%%%%%%%%%%%%%%%%%%%%%%%%%%%%%
\subsection{The $v_{n}$ measurements}

\begin{figure}[ht]
\centering
\includegraphics[width=130mm,angle=270]{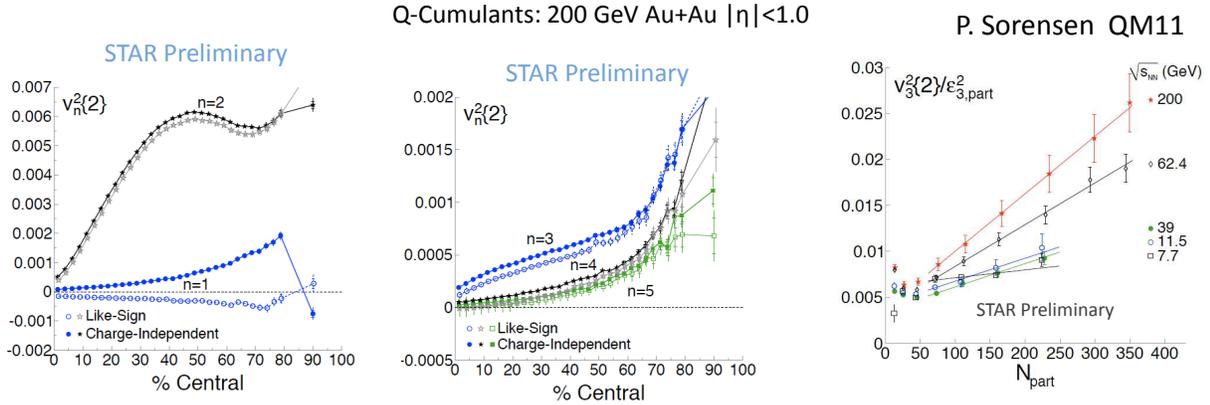}

\vspace{-65mm}

\caption{STAR $v^2_{n} (2)$ measurements as a function of centrality and collision energy.} 
\label{figure_paul_sorensen_qm11_v3}
\end{figure}

The higher order $v_{n}$s, especially those odd order items, has been noticed in the recent years as the possible explanations of the ``cone'' and ``ridge'' structures. In Figure.\ref{figure_paul_sorensen_qm11_v3} the recent STAR $v^2_{n} (2)$ measurements are shown. ~\cite{paul_sorensen_qm11_proceeding} Here the $v_{3}$ items exhibit effects of elliptic overlap geometry and follow an $N_{part}\epsilon^{2}_{3,part}$ trend similar to $v_{2}$ ~\cite{q_cumulants_paper}, indicating the possible source of initial density fluctuations manifesting into momentum space ~\cite{ma_wang_vn_paper}. 

\begin{figure}[ht]
\centering
\includegraphics[width=130mm,angle=270]{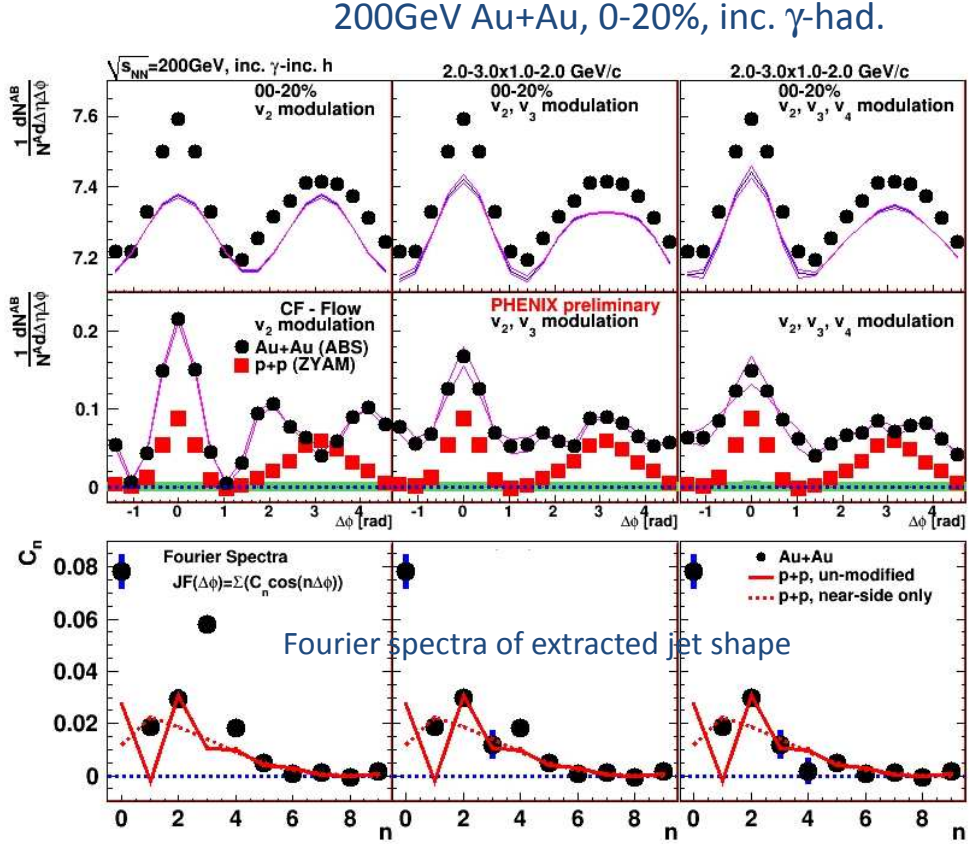}
\caption{PHENIX central Au+Au correlations, with different orders of $v_{n}$ subtracted.} 
\label{figure_shinichi_esumi_qm11_v3}
\end{figure}

Since such fluctuation models agree with data, showing at most central collisions the $v_{n}$ drop with n at low $p_{T}$ but $v_{3} ~ v_{2}$ at intermediate $p_{T}$, it is natural to ask if these $v_{n}$ together can reproduce the ``cone'' and ``ridge'' structures at RHIC. In Figure.\ref{figure_shinichi_esumi_qm11_v3} it is proved to be a great success in central Au+Au data, where the ``mach-cone'' is almost gone ~\cite{shinichi_esumi_qm11_proceeding}, although remaining medium effect can possibly exist.

However, this is not the end of story yet. Since the 2nd and 3rd order event planes, $\Psi_{2}^{EP}$ and $\Psi_{3}^{EP}$ respectively, are weakly correlated ~\cite{shinichi_esumi_qm11_proceeding}, it means the measurement that $v_{2}$ modulations subtracted correlation shapes still keep strong $\Psi_{2}^{EP}$ dependence can not be explained by pure $v_{3}$ ~\cite{fuqiang_wang_corr_ep_paper}. Currently the measured $v_{n}$s have weak $|\eta|$ dependence from long $\Delta\eta$ away $\Psi_{2}^{EP}$ and $\Psi_{3}^{EP}$, but the long $\Delta\eta$ non-flow contributions can't be excluded yet. Meanwhile, the higher order $v_{n}$ contributions may also be needed even if their magnitude is small.

\begin{figure}[ht]
\centering
\includegraphics[width=130mm,angle=270]{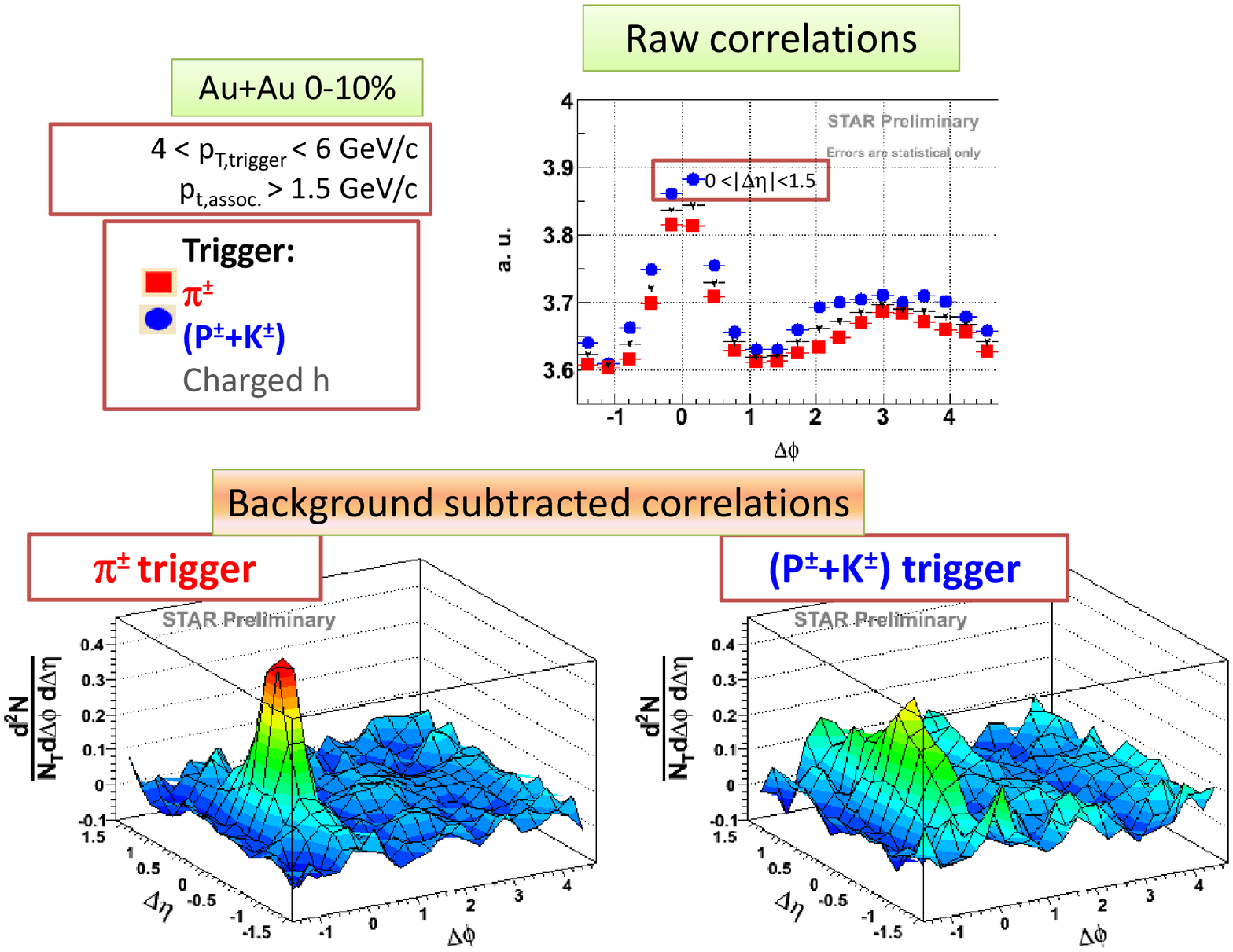}
\caption{STAR central Au+Au correlations, with different PID-ed triggers. The $\Delta\phi$ projections show mass-splitting effect on ``cone'' structure, while the $\Delta\eta$ projections show that on ``ridge''.} 
\label{figure_kolja_kauder_qm11_pid_correlation}
\end{figure}

The next step of studying $v_{n}$ modulations will be the PID-ed measurements, similar to what was done in the single particle $R_{AA}$. The $kE_{T}$ and $n_{q}$ scaling $v_{2}$ has long be used as an evidence of partonic flow within QGP ~\cite{phenix_pid_v2_ncq}, and the recent PID-ed $v_{3}$ measurements at RHIC are consistent with this ncq-scaling picture at intermediate $p_{T}$ region ~\cite{roy_lacey_qm11_proceeding}. Thus the correlation functions are expected to show an evident mass splitting effect, based on higher order $v_{n}$ modulation pattern, on structures such as ``cone'' and ``ridge''.
This is confirmed in the STAR PID-ed trigger correlation in Figure.\ref{figure_kolja_kauder_qm11_pid_correlation}, where trigger particles are grouped into charged pions and charged kaons/protons, then the low-$p_{T}$ associated particles are plotted. A clear mass splitting effect between triggers exists ~\cite{kolja_kauder_qm11_proceeding}.

The PID-ed $v_{n}$s can bring more questions. A much weaker centrality dependence of $v_{3}$ at intermediate $p_{T}$ has been observed at RHIC and LHC, contrary to that of $v_{2}$, and this is commonly considered an evidence of $v_{3}$ being caused by initial state density inhomogeneity ~\cite{phenix_v3_paper}, as were predicted by theory models. 
On the other hand, the baryons ``anomaly'' enhancements relative to mesons in A+A collisions are well-known to be centrality dependent.
If $v_{3}$ is partonic flow as indicated from RHIC/LHC data, then the weak centrality dependence of $v_{3}$ has to be due to a complex convolution among $\epsilon^{2}_{3,part}$, PID-ed $v_{3}$, and baryon anomaly. Is it a simple coincidence, or an indication of deeper relations between these physics mechanisms?
This is being studied through multiple analysis working in progress, including the $v_{3}$ modulations in the PID-ed trigger correlations, and non-flow effect in long $\Delta\eta$ correlations.

%%%%%%%%%%%%%%%%%%%%%%%%%%%%%%%%%%
\subsection{High-$p_{T}$ triggers and jets correlations}

\begin{figure}[ht]
\centering
\includegraphics[width=150mm]{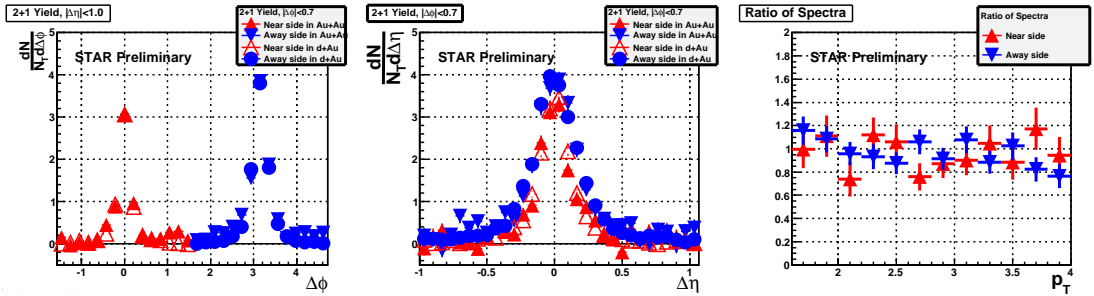}
\caption{STAR 2+1 correlations in d+Au and central Au+Au, with back-to-back high-$p_{T}$ triggers. The $\Delta\phi$ projections and $\Delta\eta$ projections both show similar shapes and magnitudes between d+Au and central Au+Au data after $v_{2}$ modulation subtracted, indicating no evidence of $v_{3}$. The ratios of associated spectra also show no evident modification.} 
\label{figure_hua_pei_2plus1_correlation}
\end{figure}

While the low-$p_{T}$ $v_{n}$ are mainly from collective effect, the high-$p_{T}$ particles measured at RHIC are dominated by jet source. Since the high-$p_{T}$ $v_{2}$ isn't approaching zero ~\cite{phenix_highpt_pi0_v2}, this is a strong evidence of jet quenching in medium. Then a similar question rises, do jets also induce $v_{3}$?
This is studied from more than one ways. 
First, a pair of high-$p_{T}$ triggers are used as proxies of jets, called ``2+1'' correlations. On contrary to the normal 2-particle correlations, the two triggers are required to be back-to-back in the azimuthal plane to tag ``back-to-back hard-scattering'' events. The low-$p_{T}$ associates are then studied around both triggers with $v_{2}$ modulation subtracted. The energy asymmetry between the two triggers are also varied as a method to control the relative medium travelling length of both partons ~\cite{star_2plus1_paper, hua_pei_wwnd11_proceeding}. In Figure.\ref{figure_hua_pei_2plus1_correlation} a set of typical 2+1 correlation functions are shown, and no $v_{3}$ observed.

\begin{figure}[ht]
\centering
\includegraphics[width=130mm,angle=270]{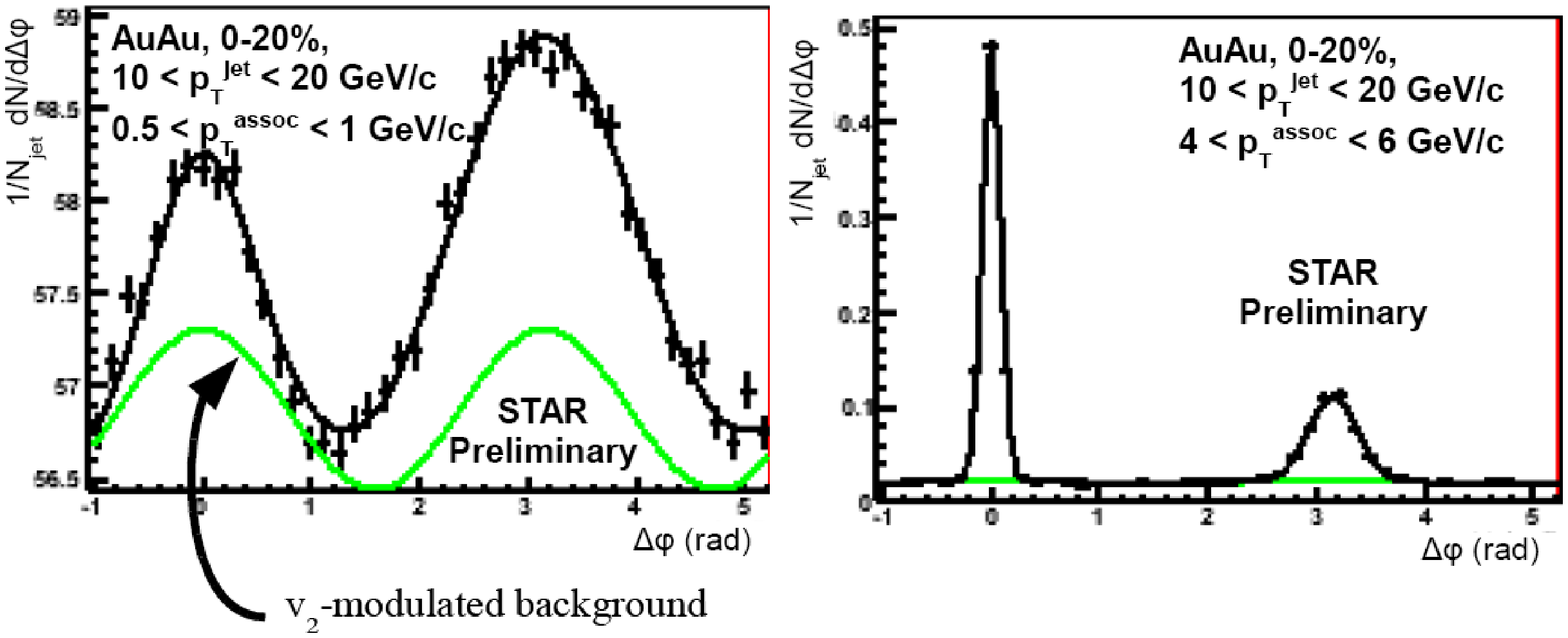}

\vspace{-60mm}

\caption{STAR jet-hadron correlations in central Au+Au. The $\Delta\phi$ projections are plotted on both low- and high-$p_{T}$ associates. The dashed lines show current estimation of $v_{2}$ modulations.} 
\label{figure_alice_ohlson_jet_correlation}
\end{figure}

Second, the fully reconstructed jets are directly applied as triggers. In Figure.\ref{figure_alice_ohlson_jet_correlation} the axis of jets are used as trigger direction and all particles are plotted around this axis. While the $v_{2}$ modulations are evident especially at low-$p_{T}$ region as expected, the $v_{2}$ subtracted correlation functions don't leave much space for possible higher-order $v_{n}$ such as $v_{3}$ ~\cite{alice_ohlson_qm11_proceeding}.

%%%%%%%%%%%%%%%%%%%%%%%%%%%%%%%%%%
\section{Cold nuclear matter effect}

\begin{figure}[ht]
\centering
\includegraphics[width=130mm,angle=270]{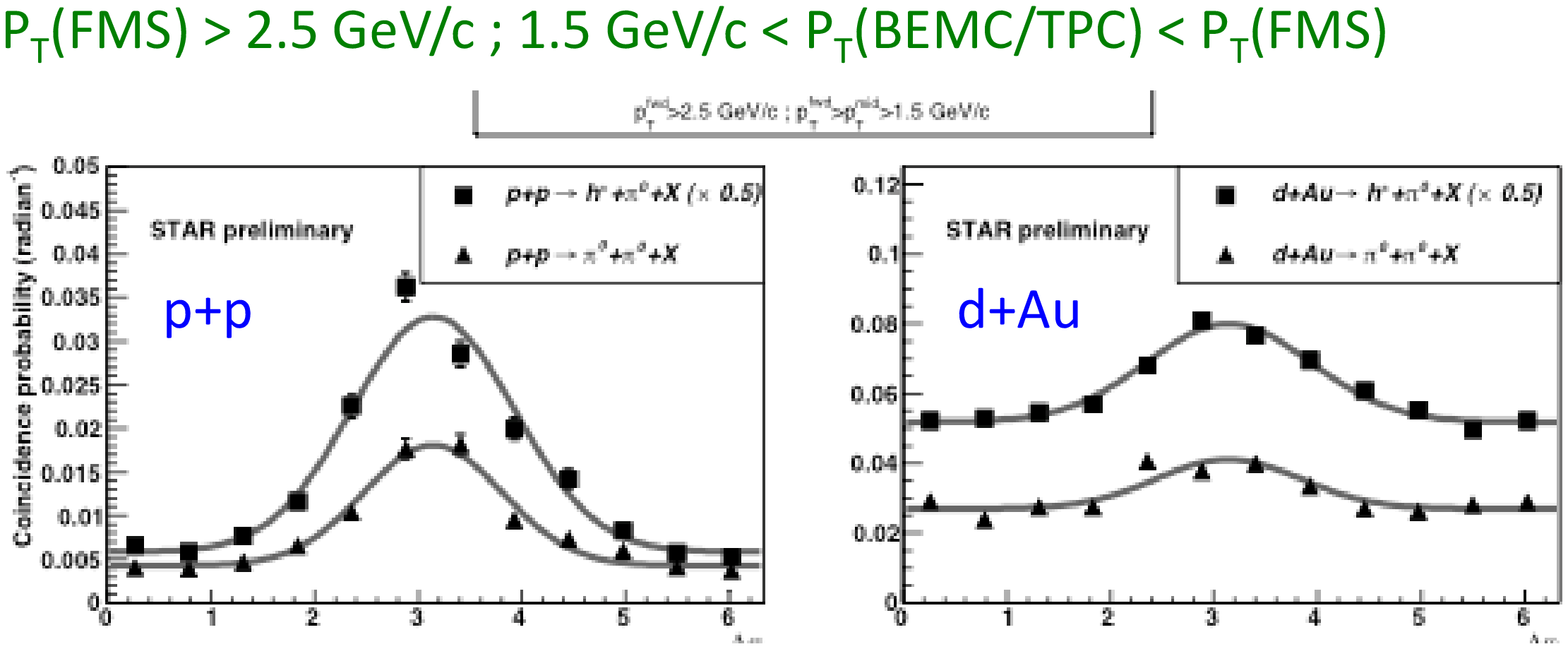}

\vspace{-60mm}

\caption{STAR forward-central correlations in p+p and d+Au.} 
\label{figure_ermes_braidot_dau_correlation}
\end{figure}

It is important to understand that all the medium properties from correlation measurements, either $v_{n}$ or jet correlations, are induced by hot nuclear matter effects in addition to the modification of nuclear target relative to ``vacuum'' collisions of protons and DIS collisions.
Therefore, the cold-nuclear-matter (CNM) effect has to be studied as a baseline. 
The Figure.\ref{figure_ermes_braidot_dau_correlation} shows the correlations in d+Au comparing to p+p. The triggers and associates are selected to be separated in broad $\Delta\eta$ similar as those $v_{n}$ measured in event-plane method. Currently, no significant broadening or away-side peak suppression observed in d+Au comparing to p+p ~\cite{ermes_braidot_dau_correlation_thesis}. 
This proves at current stage that the observed long $\Delta\eta$ higher-order $v_{n}$ are still hot nuclear matter specific.

%%%%%%%%%%%%%%%%%%%%%%%%%%%%%%%%%%
\section{Summary and outlook}

Recent measurements of higher order Fourier harmonics $v_{n}$ at RHIC has been an important role to disentangle different sources of physics, by successfully reproducing the correlation structures (cone/ridge) with little help from jets-medium interaction.
While these higher-order $v_{n}$ are widely considered to be produced by initial geometry fluctuations, more quantitative analysis and theory predictions are necessary, including their dependence on $p_{T}$, $\Delta\eta$, centrality, PID, etc.
At high-$p_{T}$ end, the hadron correlations using multiple high-energy triggers and/or fully reconstructed jets show no signal of higher-order $v_{n}$ on contrary to $v_{2}$. This observation also supports the assumptions of $v_{2}$ and high-order $v_{n}$ from different sources.
Meanwhile, no evident cold-nuclear-matter effect was observed for higher-order $v_{n}$, indicating they are hot nuclear matter specific.

\bigskip % extra skip inserted

\end{document}